\documentclass[]{spie}  
\pdfoutput=1

\usepackage{amsmath,amsfonts,amssymb}
\usepackage{graphicx}
\usepackage[colorlinks=true, allcolors=blue]{hyperref}

\title{Design and Fabrication of Metamaterial Anti-Reflection Coatings for the Simons Observatory}

\author[a]{Joseph E. Golec}
\author[a,b,c,d]{Jeffrey J. McMahon}
\author[e]{Aamir M. Ali}
\author[a]{Grace E. Chesmore}
\author[f]{Leah Cooperrider}
\author[g]{Simon Dicker}
\author[h]{Nicholas Galitzki}
\author[b]{Kathleen Harrington}
\author[b]{Rebecca Jackson}
\author[e]{Benjamin Westbrook}
\author[i]{Edward J. Wollack}
\author[g]{Zhilei Xu}
\author[g]{Ningfeng Zhu}
\affil[a]{Department of Physics, University of Chicago, Chicago, IL, USA}
\affil[b]{Department of Astronomy and Astrophysics, University of Chicago, Chicago, IL, USA}
\affil[c]{Kavli Institute of Cosmological Physics, University of Chicago, Chicago, IL, USA}
\affil[d]{Enrico Fermi Institute, University of Chicago, Chicago, IL, USA}
\affil[e]{Department of Physics, University of California-Berkeley, Berkeley, CA, USA}
\affil[f]{Department of Physics, University of Michigan, Ann Arbor, MI, USA}
\affil[g]{Department of Physics and Astronomy, University of Pennsylvania, Philadelphia, PA, USA}
\affil[h]{Department of Physics, University of California-San Diego, La Jolla, CA, USA}
\affil[i]{NASA/Goddard Space Flight Center, Greenbelt, MD, USA}

\authorinfo{Further author information: (Send correspondence to J.E.G)\\E-mail: golecjoe@uchicago.edu}

\begin{document} 
\maketitle

\begin{abstract}

The Simons Observatory (SO) will be a cosmic microwave background (CMB) survey experiment with three small-aperture telescopes and one large-aperture telescope, which will observe from the Atacama Desert in Chile. In total, SO will field over 60,000 transition-edge sensor (TES) bolometers in six spectral bands centered between 27 and 280 GHz in order to achieve the sensitivity necessary to measure or constrain numerous cosmological quantities, as outlined in The Simons Observatory Collaboration et al. (2019). These telescopes require 33 highly transparent, large aperture, refracting optics. To this end, we developed mechanically robust, highly efficient, metamaterial anti-reflection (AR) coatings with octave bandwidth coverage for silicon optics up to 46 cm in diameter for the 22-55, 75-165, and 190-310 GHz bands. We detail the design, the manufacturing approach to fabricate the SO lenses, their performance, and possible extensions of metamaterial AR coatings to optical elements made of harder materials such as alumina.
\end{abstract}


\section{INTRODUCTION}

\noindent The Simons Observatory (SO) is an up-coming ground-based survey experiment that will provide the most sensitive measurements of the cosmic microwave background to date in order to constrain fundamental cosmological properties of our universe \cite{Ade19, Galitzki18}. To make high fidelity measurements of the CMB, SO will use silicon refractive optics to focus light onto detector arrays. Silicon is an excellent choice of lens material for the millimeter and sub-millimeter wavelengths due to its low loss and high index of refraction which leads to high-throughput and large field of view optical designs ideal for a large sky CMB survey. However, the high index of refraction of the silicon lenses also means that a significant fraction of the incident light is reflected. This not only causes less light to reach the detectors, thus causing a decrease in overall sensitivity, it can lead to other non-ideal instrument sytematics due to multiple reflections in the instrument. The undesirable consequences of the high index means that the lenses must have an anti-reflection (AR) coatings in order to deliver state-of-the-art measurements of the CMB.

The standard method to AR coat lenses is to layer thin films, usually made of a plastic material, onto the lens surface. The thickness and index of the thin films can be tuned to optimize the reflection across a given frequency band. While this works in many applications, the lenses for SO will be kept at cryogenic temperatures and any plastic AR coating risks delamination from the lens due to a mismatch between the coefficient of thermal expansion between silicon and plastic. To solve this problem metamaterial AR coatings have been implemented in CMB experiments to great success \cite{Datta13, Coughlin18,10.1117/12.2233125}. Metamaterial AR coatings consist of sub-wavelength features either placed onto or machined into an optical surface. Those sub-wavelength features then act as effective dielectric layers akin to traditional thin film plastic coatings. Since the metamaterial coatings are made of the lens material itself there is no risk of delamination of the AR coating from the optic. The shape and dimensions of the sub-wavelength features of the metamaterial coating can be tuned to result in sub-percent reflections across octave bandwidths which is optimal for experiments like SO.

We present the work done to realize metamaterial AR coatings for lenses in the three SO observing bands and at the necessary production scale for an experiment the size of SO. The organization of this paper is as follows: Section 2 presents the design of the AR coatings for all three SO observing bands. Section 3 describes the production process of the AR coatings and presents the achieved production rates. In Section 4 the results of reflection measurements taken of the AR coatings are presented. Section 5 discusses the possible extensions of this AR coating technique to alumina, another material used for optical elements in CMB observation. Finally, Section 6 concludes with discussion of the how this AR coating technology fits into the broader context of the sensitivity of SO and comments on the feasibility of metamaterial AR coatings for future CMB missions at a scale larger than SO.

\section{Design}

\noindent Metamaterial AR coatings consisting of sub-wavelength features have been used to mitigate reflections off optical elements for many years. In general, the idea behind metamaterial AR coatings is to create a periodic array of features that are sufficiently smaller than the wavelength of the incident light. By tuning the geometries of those features, reflections can be minimized across a given bandwidth. The geometry and fabrication of the sub-wavelength features varies greatly depending on the application and the wavelength of the incident light. Raut et al (2011) gives a general review of designs and fabrication techniques used to create AR coatings \cite{C1EE01297E}.     

\begin{figure} [ht]
   \begin{center}
   \begin{tabular}{c} 
   \includegraphics[height=5cm]{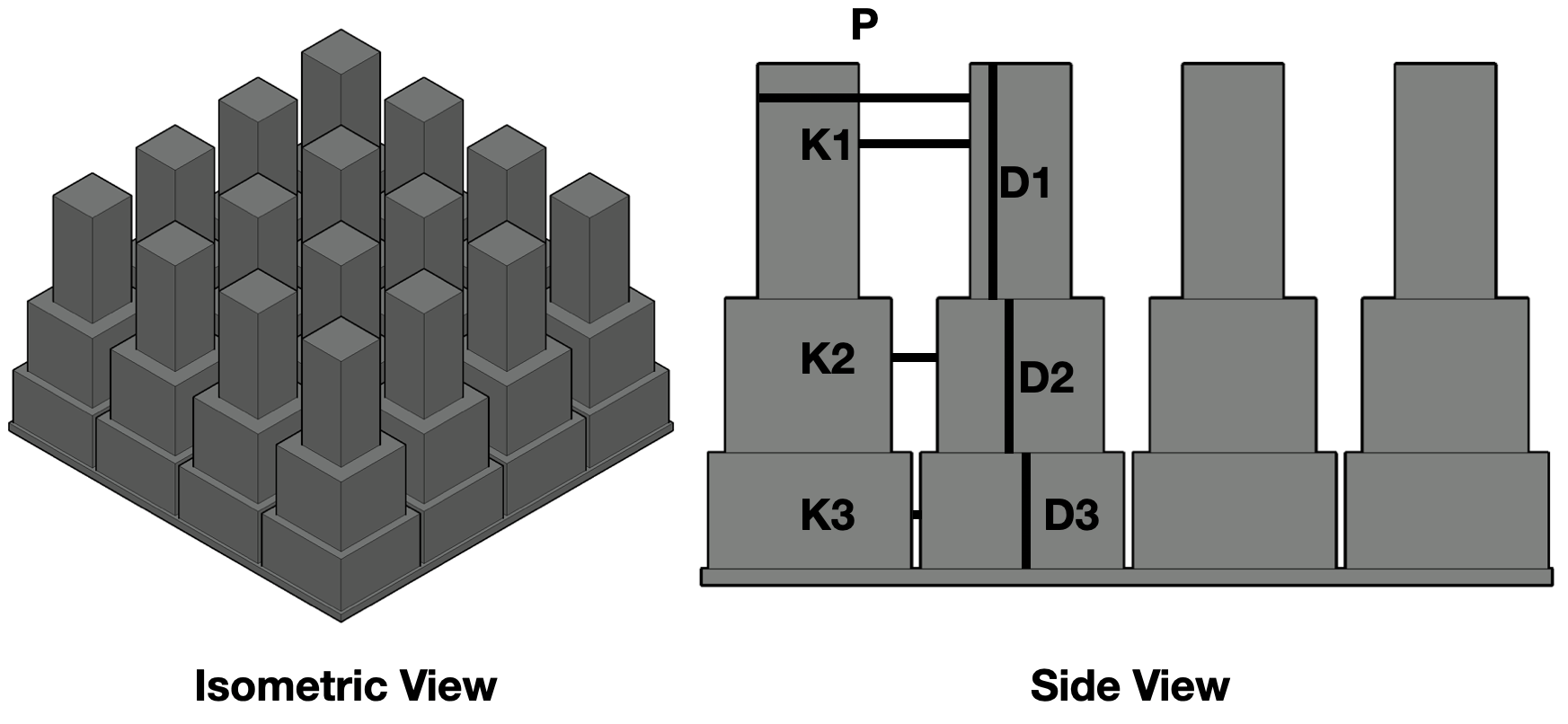}
   \end{tabular}
   \end{center}
   \caption[example] 
   { \label{fig:design} 
(Left) Isometric view of a fiducial three-layer AR coating design. (Right) Side view of the fiducial three-layer design with the relevant design parameters labeled.}
\end{figure}

\noindent The design of the sub-wavelength features we present here closely follows the geometry presented in Datta et al (2013). \cite{Datta13} The features consist of metamaterial layers that are an array of square ``stepped-pyramids". This design was chosen because these features are easily fabricated with a silicon dicing saw, where the saw makes a series of nested cuts across an optic, the optic is then rotated 90 degrees, and the series of cuts are made again. Figure \ref{fig:design} shows a fiducial model of a three-layer metamaterial AR coating. In principle, the number of layers can be increased to accommodate large bandwidth coverage, but this is subject to physical constraints such as dicing blade thickness and aspect ratio. Metamaterial coatings with five-layers have been demonstrated and show excellent performance over more than an octave bandwidth \cite{Coughlin18}. Following the fiducial design the pitch, or the spacing between the periodic cuts, each layer's width (kerf), and depth must be optimized to minimize the reflections across the observing bands.

The SO will observe in three dichroic frequency bands; the low frequency (LF), mid frequency (MF), and ultra-high frequency (UHF). The band edges for those three channels observe from 23-47 GHz, 70-170 GHz, and 195-310 GHz for the respective LF, MF, and UHF bands. We begin the optimization process by modeling the physical AR coating structure in the finite-element analysis software, Ansys High Frequency Structure Simulator (HFSS)\footnote{https://www.ansys.com/products/electronics/ansys-hfss}. Rather than start the optimization process from scratch we began the optimization process with the three-layer 75-170 GHz metamaterial AR coating presented in Coughlin et al (2018). 

The pitch of the sub-wavelength features is set by the criterion for diffraction which is given by Equation 1 in Datta et al (2013)\cite{Datta13}

\begin{align}
p < \frac{\lambda}{n_{\text{si}}+n_{\text{i}}\sin \theta_i}
\end{align}
Where p is the pitch, $\lambda$ is the wavelength corresponding to the upper edge of the frequency band, $n_{\text{si}}$ and  $n_{\text{i}}$ are the indices of refraction of silicon and the incident medium (in this case vacuum) respectively, and $\theta_i$ is the angle of incidence of light on the surface of the lens. In our design we choose the pitch such that this criterion is met up to the upper edge of the observing frequency band at an angle of 15 degrees which is roughly the average angle of incidence of a light ray in the telescopes. 

With the pitch set we then optimize the AR coating design using two free parameters, the kerf and depth, per each metamaterial layer. Therefore, for a three layer coating there are six parameters to optimize and for a two layer coating there are four. An optimization algorithm in HFSS is used to vary the kerfs and depths of the metamaterial layers to achieve the lowest reflection possible across the SO MF band. Since the size of the sub-wavelength features dictate at what frequencies the coating is effective, the dimensions of the parameters of the AR coating can be scaled up or down to cover the LF and UHF bands. However, that scaling may not produce an optimized AR coatings in that new frequency band or the resulting design may not be physically realizable with dicing saw blades. 

For the SO LF band, the optimized MF AR coating design parameters were scaled up and then re-optimized. The resulting coating from that optimization achieved sub-percent reflection across the LF band. Finally, the MF AR coating was scaled down to try and cover the UHF band, but kerf of the third layer became thinner than any physically producible dicing saw blade. This drove the design of the UHF coating from a three-layer design to a two-layer design. After optimizing the two-layer UHF design we still achieved a simulated sub-percent reflection across the band due to the UHF's more narrow fractional bandpass. 

After the completed optimization we then have three AR coating designs that all achieve sub-percent reflection across their respective frequency bands. The dimensions of the three optimized AR coatings for the LF, MF, and UHF bands are presented in Table \ref{tab:DesignParams}. The parameters refer to the Figure \ref{fig:design}. Note that the UHF design is a two-layer and therefore the dimensions of the third layer are non-applicable. The simulated performance of the AR coatings at normal incidence are presented later in Section \ref{sec:performance}.

\begin{table}[ht]
\caption{Parameters of the three AR coating designs} 
\label{tab:DesignParams}
\begin{center}       
\begin{tabular}{|l|l|l|l|}
\hline
\rule[-1ex]{0pt}{3.5ex}   & LF & MF & UHF  \\
\hline
\rule[-1ex]{0pt}{3.5ex}  Pitch (P) & 1.375 mm  & 0.450 mm  &  0.266 mm \\
\hline
\rule[-1ex]{0pt}{3.5ex}  Kerf 1 (K1) & 0.735 mm  & 0.245 mm  & 0.122 mm   \\
\hline
\rule[-1ex]{0pt}{3.5ex}  Depth 1 (D1) & 1.520 mm  & 0.452 mm  & 0.200 mm  \\
\hline
\rule[-1ex]{0pt}{3.5ex}  Kerf 2 (K2) & 0.310 mm  & 0.110 mm  & 0.033 mm \\
\hline
\rule[-1ex]{0pt}{3.5ex}  Depth 2 (D2) & 1.000   & 0.294 mm  & 0.120 mm \\
\hline
\rule[-1ex]{0pt}{3.5ex}  Kerf 3 (K3) & 0.070 mm  & 0.025 mm  & - \\
\hline 
\rule[-1ex]{0pt}{3.5ex}  Depth 3 (D3) & 0.750 mm  & 0.234 mm  & - \\
\hline
\end{tabular}
\end{center}
\end{table}

\section{Production}

\noindent The SO will deploy over 30 silicon lenses which is the most by any single experiment to date and therefore the production rate of the metamaterial AR coatings for those lenses must be high enough to follow the deployment timeline. This combined with the added complications that we need to dice the largest diameter silicon lenses deployed on an experiment to date and the complex surface profiles of some of the lenses \cite{Ali2020}  lead to the development of a custom silicon dicing saw system. 

\begin{figure} [ht]
   \begin{center}
   \begin{tabular}{c} 
   \includegraphics[height=7cm]{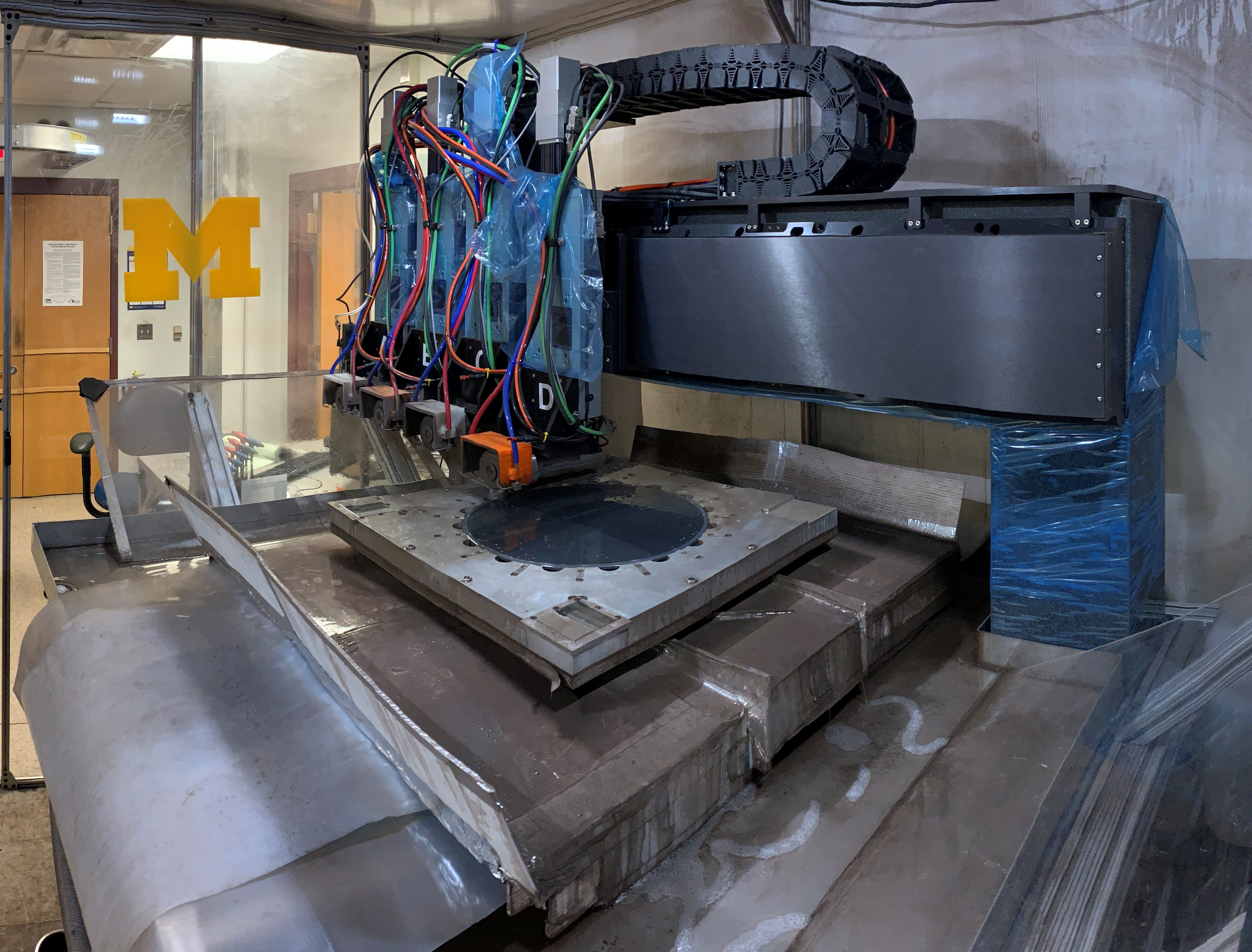}
   \end{tabular}
   \end{center}
   \caption[example] 
   { \label{fig:saw} 
Picture of the dicing saw system.}
\end{figure}

The saw system uses nickel alloy dicing saw blades embedded with diamonds to dice the metamaterial features into the lens surface. Figure \ref{fig:saw} shows a picture of the dicing saw system cutting a lens surface. There are numerous features that allows for significant increases in production rate compared to previous efforts to produce metamaterial AR coatings for CMB experiments. First there are multiple dicing spindles which can each be fit with a different blade corresponding to different cuts in the AR coating's design. By mounting all of the blades at once we can AR coat an entire optical surface without changing any blades. This provides for nearly continuous operation of the saw system. Another feature of the custom system is that the lens is mounted to a rotary stage which allows for the lens to remain mounted to the system for the entirety of the fabrication process. This eliminates the need to perform metrology on the cut lens surface which is a time consuming process. Careful commissioning and calibration of this dicing saw system lead to micron accurate stage positioning and repeatability which is well within the tolerances required for AR coating application presented here.

The general production procedure of the AR coating on a lens is described hereafter. A lens is mounted to the dicing saw and metrology is taken of its surface using a sub-micron accurate metrology probe mounted to the system. A program then takes the surface metrology, fits a model surface to the data, and generates program files that are used to command the system to dice the cuts. The room that the dicing saw is situated in is temperature regulated and the dicing process uses temperature controlled flood cooling. This temperature regulation is necessary to ensure the lens surface does not thermally expand or contract during the cutting process. The design-specific blades are then mounted to the spindles, and are ``prepared," to eliminate diamond burs on the blade and to ensure the blade is circular. This is done by making several cuts in a special dressing block made to hone dicing saw blades. Test cuts are then diced into a small silicon wafer affixed to the side of the mounted lens. These cuts are then inspected and their dimensions measured with a microscope. This is a check that all the blades are dicing properly and the CNC system is correctly programmed to dice the cuts into the lens. The layers of the AR coating are diced into the lens, one at a time, from the largest to the smallest cut. After the layers are diced into the lens, it is then rotated 90 degrees and the process is repeated to fully realize the AR coating. After the AR coating is completely diced, additional test cuts are made in the sacrificial wafer to monitor if any cutting abnormalities may have arisen during the fabrication. After one optical surface of a lens is finished it is flipped and the procedure repeated on the other side. After both sides of a lens have been AR coated, the lens is cleaned with water in an ultrasonic cleaning bath.

That is the process for the MF and the UHF coatings but for longer wavelengths where the feature size is much larger we must modify this approach. Dicing blades cannot be fabricated to have an arbitrary kerf, so for the top two layers of the LF coating we use three defining cuts and two clearing cuts to create a kerf that is much wider than the maximum blade thickness. In order to not load the dicing blades with too much cutting force we make multiple passes of defining and clearing cuts to realize the full depth of the top two LF layers.  

\begin{figure} [t]
   \begin{center}
   \begin{tabular}{c} 
   \includegraphics[height=7cm]{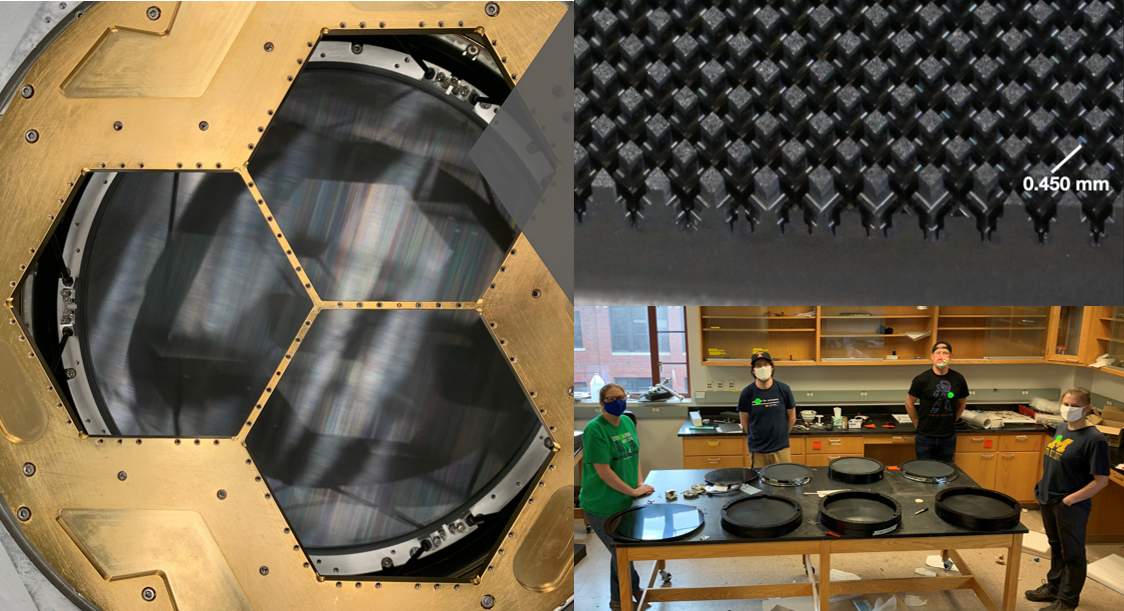}
   \end{tabular}
   \end{center}
   \caption[example] 
   { \label{fig:lens} 
(Left) Picture of a SO LAT lens installed in an optics tube. (Right Top) A zoomed in image of the MF metamaterial AR coating. (Right Bottom) A Picture of the production team with six SO lenses. The three closest to the camera are a set of SO Small Aperture Telesscope lenses and the three farther away are a set of SO Large Aperture Telescope lenses.}
\end{figure}

In total we have fabricated nine lenses to date for SO. All nine were coated with the MF coating. Figure \ref{fig:lens} shows an image of one of the MF lenses installed inside an optics tube (see paper \textbf{\#11453-183} in these proceedings for a discussion of the SO optics tubes). It also shows a zoomed in picture of the fabricated coating. In addition to the MF lenses for SO, three LF lenses using the SO design were fabricated for the AdvACT experiment. The UHF coating has yet to be fabricated. At the end of the fabrication run for the MF lenses we achieved a production rate of one lens per week. The defect rate was around 100 broken pyramid features out of a million which is not expected to impact the lens quality or the AR coating performance.    

\section{Optical Performance}
\label{sec:performance}

\begin{figure} [ht]
   \begin{center}
   \begin{tabular}{c} 
   \includegraphics[width=0.75\linewidth]{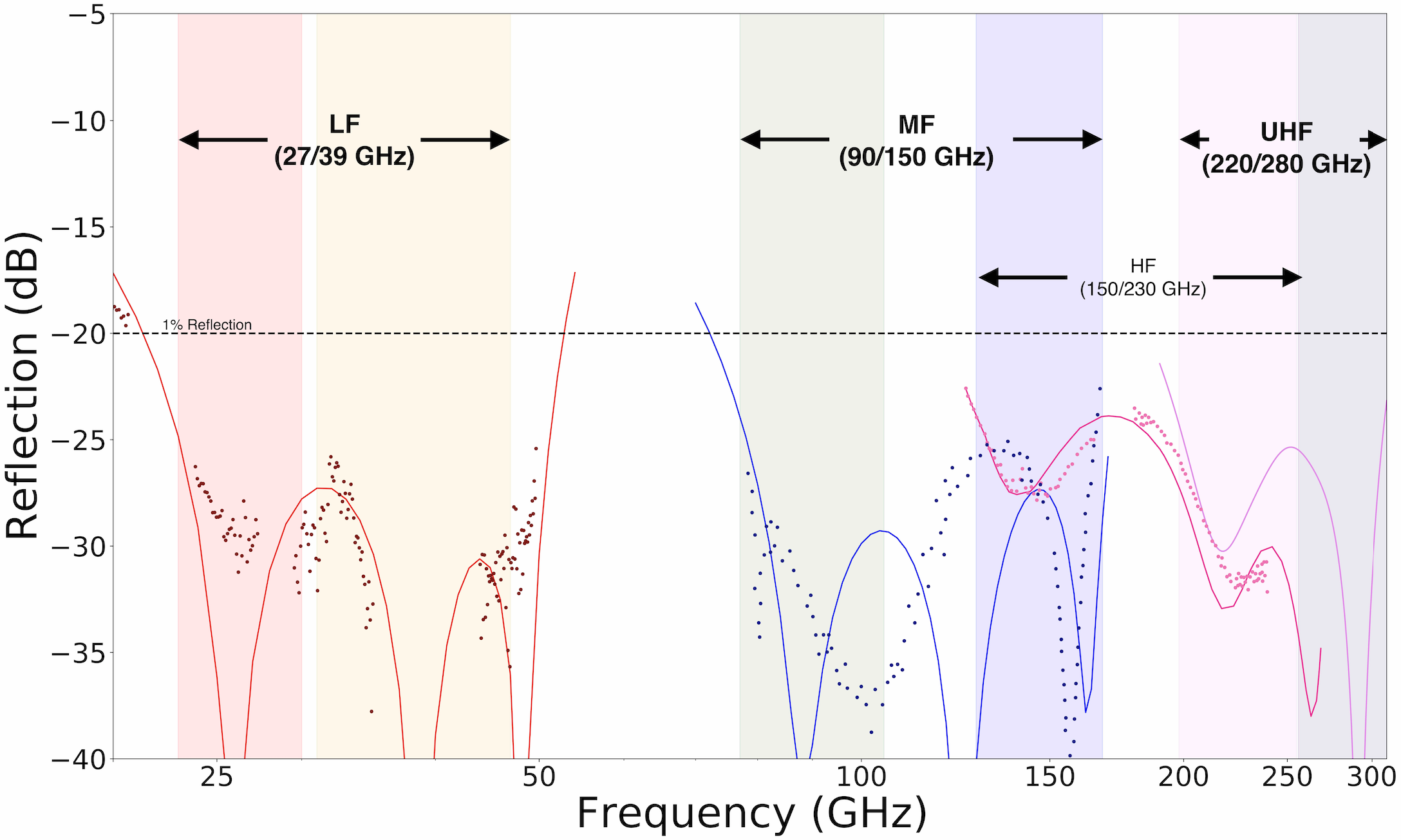}
   \end{tabular}
   \end{center}
   \caption[example] 
   { \label{fig:reflectionplot} 
Plot of the reflection performance of the SO AR coatings. The solid line represent the simulated performance of the AR coating and the dots represent measurements.}
\end{figure}

\noindent The lenses were tested and the optical performance measured using a coherent reflectometer. The reflectometer setup is described in detail in Chesmore et al.(2018) \cite{Chesmore18}. The lenses are mounted like in Figure 1 of the Chesmore paper with the flat side down toward the parabolic mirrors. In cases where the lens does not have a flat side, we measure the concave side as close to the center of the lens as possible where it is the most flat. The results of the measurements is summarized in Figure \ref{fig:reflectionplot}. The presented data for the LF AR coating is from lenses made for the AdvACTPol experiment which share the same design and fabrication procedure as the SO lenses. This data shows good agreement with simulations with sub-percent reflections across the LF bands. Due to the coronavirus pandemic, it was not possible to make reflection measurements of the MF AR coatings produced for SO. The data for the MF coating presented in Figure \ref{fig:reflectionplot} is of the AR coatings produced for the ACTPol experiment which have a slightly different design from the SO coating. The performance of all of the measured coatings so far have achieved sub-percent reflection across their bands. Since the UHF coating has yet to be fabricated we have included the simulation and performance of the high frequency (HF) metamaterial AR coating used for the AdvACT experiment to show that sub-percent reflection is achievable at frequencies comparable to the UHF frequencies.

\section{Extensions to Alumina}

\noindent The success of metameterial AR coatings at achieving sub-percent reflection across observing bands in silicon motivates investigating if this method can be extended to alumina, another material used for millimeter-wave optics. Alumina is used as an IR blocking filter in SO and, like silicon, has a relatively high index of refraction, so AR coating is just as important for the alumina optical elements. Current methods of AR coating alumina optics are to glue layers of epoxies and plastics on the surface \cite{Rosen:13}. While this results in an effective AR coating, it still suffers from the differential thermal contraction between the AR coatings and the optic, which can lead to delamination of the AR coating. This failure may not occur on early thermal cycles of the optic, but over the course of subsequent observing campaigns where the optics are cryogenically cycled numerous times there is no guarantee that the AR coating will remain affixed to the optic. While considerable effort has been made to reduce or prevent the delamination of the plastic coatings through careful surface preparation and laser strain relief, metamaterial AR coatings avoid this differential thermal contraction all-together.

While it is straightforward to come up with a design of a metamaterial AR coating for alumina the fabrication of that coating is not straightforward due to alumina's hardness. Alumina shares the same chemical composition of sapphire but is not in a crystalline form. Instead alumina optics are made by taking aluminum oxide power and binding it together with heat and pressure in a mold. The resulting optic is nearly as hard as sapphire, which makes machining possible but much more difficult than machining silicon, meaning issues like blade wear become an issue.

To overcome the difficulty of dicing alumina, we began testing different dicing blade types and have found that a combination of resin and nickel-alloy blades, each with different diamond grit and density, can be used to fabricate a diced metamaterial coating in alumina. Tool wear is still an issue with these more resilient dicing blades however we have found that the tool wear scales linearly with the amount of material cut so it can be compensated for in the saw cutting software. 

\begin{figure} [ht]
   \begin{center}
   \begin{tabular}{c} 
   \includegraphics[height=7.5cm]{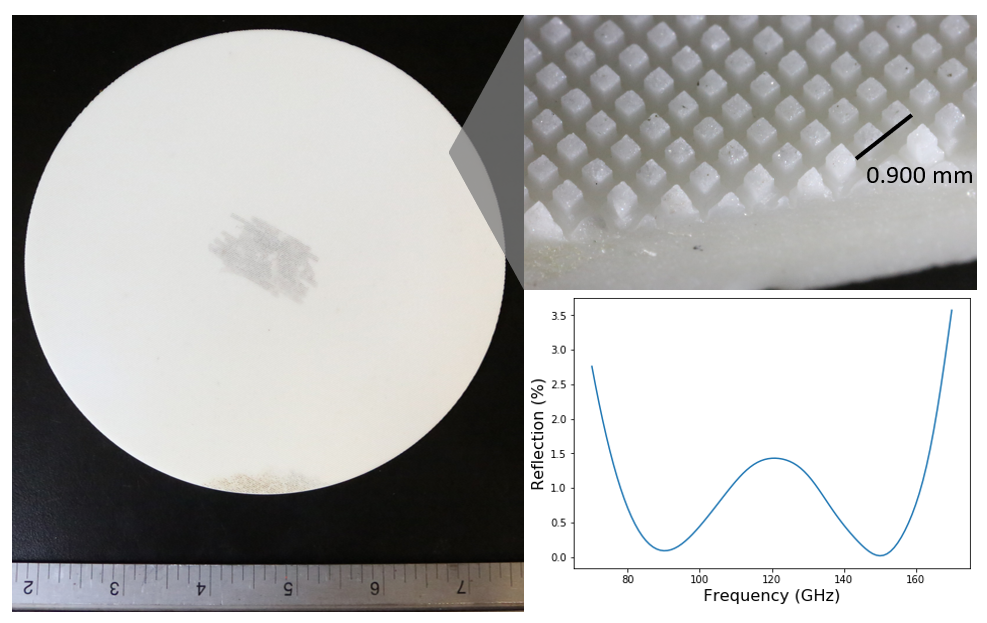}
   \end{tabular}
   \end{center}
   \caption[example] 
   { \label{fig:alumina} 
(Left) Picture of the six-inch alumina wafer coated with a prototype metamaterial AR coating. The black mark at the center is permanent marker from the fabrication process. (Right Upper) A zoomed in picture of the AR coating. (Right Lower) Plot of the simulated performance of the AR coating.}
\end{figure}

We successfully fabricated a prototype metamaterial AR coating on a six-inch diameter alumina wafer (Figure \ref{fig:alumina} Left). Due to blade thickness limitations, the alumina AR coating is only a two-layer AR coating which leads to percent-level reflections across the MF band. At time of writing we are currently measuring the performance of this coating.   

\section{Conclusions}

We have presented the design, fabrication process, and performance of the metamaterial AR coatings for the three SO bands as well as a prototype alumina AR coating for the MF band. All of these coatings achieve percent or sub-percent levels of reflection which permit sensitive and precise measurement of the CMB. In addition, we have shown that these AR coatings can be fabricated on a one to two week time scale with little to no defects. This high production rate was achieved with a custom dicing saw and is nearly the maximum rate that the coatings can be fabricated thus the limiting schedule drivers are then the procurement of the silicon on the fabrication of the non-AR coated lens blanks. Such a high production rate of the AR coatings is crucial for meeting the large demand for silicon lenses the SO imposes and reinforces the feasibility of future CMB experiments that will be at an even larger scale than SO like CMB-S4.    

\acknowledgments 
 
This work was funded by the Simons Foundation (Award \#457687, B.K.). JG is supported by a NASA Space Technology Research Fellowship (Grant 80NSSC19K1157).  ZX is supported by the Gordon and Betty Moore Foundation  

\bibliography{report} 
\bibliographystyle{spiebib} 

\end{document}